# Assessment of corticospinal tract dysfunction and disease severity in amyotrophic lateral sclerosis.

Rahul Remanan, *M.B.B.S.\**, Viktor Sukhotskiy, *Ph.D. graduate student,* Mona Shahbazi, *N.P.,*  Edward P. Furlani, *Ph.D.,*  Dale J. Lange, *M.D.*

*\* Corresponding author*

*Abstract:*

*The upper motor neuron dysfunction in amyotrophic lateral sclerosis was quantified using triple stimulation and more focal transcranial magnetic stimulation techniques that were developed to reduce recording variability. These measurements were combined with clinical and neurophysiological data to develop a novel random forest based supervised machine learning prediction model. This model was capable of predicting cross-sectional ALS disease severity as measured by the ALSFRSr scale with 97% overall accuracy and 99% precision. The machine learning model developed in this research provides a new, unique and objective diagnostic method for quantifying disease severity and identifying subtle changes in disease progression in ALS.*

Amyotrophic lateral sclerosis (ALS) is a progressively fatal neuro-degenerative condition. It is associated with a differential involvement of upper motor neuron (UMN) and lower motor neuron (LMN) pathways. To better understand the differential impact of UMN and LMN components on disease related disability and mortality, an objective quantification of different components in the ALS disease process is needed. The UMN involvement in ALS is difficult to measure objectively. Its contribution to overall disease severity is important, yet not well quantified. The most widely used clinical measures for UMN involvement are: identification of overactive reflexes, detection and grading spasticity, eliciting a plantar response and timed repetitive functional tests. Such findings are not specific and difficult to precisely quantitate.

Important components of the UMN system are the motor cortex and the corticospinal tract. The corticospinal tract is the pathway responsible for conveying kinematic (*1, 2*) and dynamic (*3*) state information from motor networks responsible for the control of voluntary muscle activity. Any functional loss in the corticomotoneuronal pathway can manifest as a loss of motor function (*1*) and impact disease severity in ALS. Hence, it is important to understand the disease severity changes in relation to corticospinal tract functional loss.

Transcranial magnetic stimulation (TMS) allows non-invasive brain stimulation of the motor cortex and is helpful to quantitate UMN involvement (*4*). Though TMS measurements are useful for measuring UMN involvement, there is an inherent variability associated with the most commonly used TMS parameters. Therefore, the results are difficult to clinically interpret. A simplified set of TMS parameters that are easier to interpret would have better utility in a clinical setting. Furthermore, these measures could provide us with meaningful insights into the extent of UMN involvement contributing to the overall disease severity. The identification of a set of TMS parameters that enable ALS disease severity prediction was the goal of this study.

*Methods:*
*Subjects and ethical review of procedures:*
An approval to perform TMS and clinical assessments of disease severity was obtained from an institutional review board. Thirty eight subjects with clinical features and diagnosis consistent with motor neuron disease (MND) were included in this study. There were no re-imbursements for participating in this study (**see supplementary methods**).

*Disease severity measures:*
The disease severity of subjects with ALS was measured using ALSFRSr(*5*) and AALS(*6*). The ALSFRSr is an easy to administer scale and a strong indicator of survival and global function (*5*). It is well-suited for identifying any possible relationship between disease severity and TMS parameters. Since, all motor evoked potentials (MEP) following TMS were recorded from abductor digiti minimi (ADM), these MEP recordings directly represent the degree of upper extremity UMN dysfunction. However, ALSFRSr lacks an objective measurement of upper extremity functional loss, a requirement for direct comparison of TMS results. The AALS on the other hand is based on a weighted score approach using a variety of subcomponent measures. Therefore, the AALS was included here as a secondary disease severity measure.

The subcomponents of AALS include: FVC, timed 20ft walking test, timed standing up from a chair, timed standing up from lying down, timed climbing up and down 4 steps of stairs, timed propelling of wheelchair 20ft, grip force testing using dynamometer, pinch grip testing using pinch gauge, Purdue peg board test, block-and-board test and manual muscle testing for both upper and lower extremities(*5*). Each of these variables has an independent predictive value for ALS disease severity. The individual measures used in this scale allowed us to dissect specific upper extremity dysfunction and a subcomponents analysis of AALS was therefore important. Also, ALSFRSr and AALS have unique characteristics as disease severity measures in ALS (*7*), hence the relevance of comparison of both measures.

Another key disease severity measure was the impact of ALS disease process on quality of life(*8*). The McGill quality of life single item scale (MQoLSiS) is an easy to administer scale that can measure ALS related quality of life changes (*9*). We independently recorded ALSFRSr, total AALS scores and MQoLSiS for disease severity, along with raw measurement values for all the individual clinical measures that constituted AALS.

***TMS and peripheral nerve stimulation:***
During each visit, the following TMS parameters were recorded: MEP amplitude, central motor conduction time (CMCT), TST amplitude ratio (rAmp) and TST area ratio (rArea). Motor threshold estimation using MEP/CMAP with a likelihood of >50% was used as a measure for cortical excitability(*10*). Latency measurements from F-wave responses were used to calculate CMCT. The motor cortex stimulation was performed using a MagPro R-30 (MagVenture Inc., Alpharetta, GA, USA) magnetic stimulator with a figure-of-eight (butterfly) geometry MC-B70 coil (MagVenture Inc., Alpharetta, GA, USA). A figure-of-eight coil has a better ability to localize the area of stimulation than a traditional circular coil as demonstrated below.

The electrical stimulations of Wrist (Wr) and erb's point (Ep) along with the corresponding electromyography (EMG) recordings were performed using an EMG console (Natus Neurology Incorporated, Middleton, WI, USA). For the Wr and the Ep electrical stimulation, adhesive 2cm diameter Ag/AgCl disposable electrodes (Natus Neurology Incorporated, Middleton, WI, USA) were used. Similar adhesive electrodes were used to record the responses from ADM muscle (**see supplementary methods**).

For maximal activation of the corticomotoneuronal pathway using the figure-of-eight coil; a biphasic, counterclockwise current direction pulse was used(*11*). The stimulation intervals for performing the study were calculated by the TST software. The three sites of stimulation for $TST_{test}$ were the motor cortex – Wr – Ep. The $TST_{test}$ responses represent the fraction of surviving motor neurons in the corticospinal tract and were the result of timed collisions dictated by the software. The $TST_{control}$ recordings were made to rule out any peripheral nerve conduction abnormalities by applying Ep – Wr – Ep sequential stimulations. The amplitude and area under the curve of the $TST_{test}$ and $TST_{control}$ responses were used to calculate the TST amplitude and area ratios (*12*).

### *TMS computational model:*

We developed three-dimensional (3D) computational models to investigate and compare the induced electric (E) field distributions generated by two different TMS coils: 1) a traditional circular coil (MagVenture C-100) and 2) the MagPro R-30 stimulator with MC-B70 figure-of-eight coil. The fields were computed using the Comsol AC/DC module with the Magnetic Fields (mef) physics interface, version 5.2 (***see supplementary methods***) (*13*). The coils were energized using a biphasic sinusoidal current. The resultant E fields for the two coil geometries subjected to this excitation are shown in ***figures 1a,b*** (***see also, supplementary methods***). This analysis shows clear differences between the field distributions of the two coil configurations. Specifically, the figure-of-eight coil produces a much more spatially focused E field than the circular coil, which is consistent with previous studies (*14*).

### *Analysis and prediction model:*

Statistical analyses were performed using R 64 bit for Microsoft Windows, version 3.3.1(*15*). A supervised machine learning (ML) model using random decision trees(*16, 17*) was built using various parameters collected during this study. The goal of this ML model was to develop an accurate predictor for ALSFRSr scores. The ML model was tested against a double-blinded ALS disease severity dataset that was collected after its development. The ML model was also compared against a multiple linear regression model (MLR) for predicting ALSFRSr scores (***see supplementary methods***).

### *Results:*
### *Clinical population:*

Following screening, a review of patient charts and the administration of informed consent, 38 subjects were included in the motor neuron disease (MND) category and underwent TMS. Applying Shapiro-Wilk test of normality on the age distribution of the subjects included in the study, the normality (p=0.60) of the sample population was established. All 38 subjects had either suspected, possible, probable, laboratory supported definite, or definite ALS according to El-Escorial criteria. The group comprised of 23 men with an average age of 56.26 (SD=11.76) years and 15 women of average age 57.87 (SD=7.69) years. The average duration of the disease for men was 2.52 (SD=2.66) years and for women, 3.80 (SD=5.03) years with no significant difference in disease duration between the two gender groups.

Among the 38 subjects, 12 had findings limited to UMN dysfunction lasting for more than 4 years. Significant bulbar dysfunction in the form of dysarthria, swallowing difficulty or respiratory problems was identified in 8 subjects. A bulbar predominant disease process, with minimal or no involvement of upper or lower extremities were observed in 4 subjects.

The average ALSFRSr score for men was 41.18 (SD=5.38) and 37.33 (SD=6.72) for women. A statistically significant difference was observed in ALSFRSr scores (p=0.01, 95% CI: 0.90, 6.80) between the two genders. The average AALS score for men was 49.86 (SD=14.94) and 55.27 (SD=19.72) for women. The mean quality of life as measured using McGill quality of life single item (MQoLSiS) questionnaire for men was 7.27 (SD=1.93) and 6.87 (SD=2.60) for women. No significant differences in AALS and MQoLSIS scores were observed between the two genders.

The average pain or discomfort associated with the neurophysiology procedures were measured using visual analog scale. The scale reported an average discomfort of 4.00 (SD=2.83) for men and 3.80 (SD=3.02) for women with no significant difference in tolerating the procedure between the two genders.

### *Triple stimulation test ratio:*

A simple linear regression was calculated to predict the dexterity task performances based on the triple stimulation test. A significant linear relationship was observed between triple stimulation test results and both the dexterity tasks used in this study. For block-and-board test based on the TST amplitude ratio (rAmp) resulted in a significant regression equation ($F(1,72)=62.82$, $R^2=0.46$, $p<0.01$). Another simple linear regression using rAmp significantly predicted the Purdue peg board results ($F(1,72)=37.16$, $R^2=0.33$, $p<0.01$). A significant linear relationship was also observed between triple stimulation test results and both of the objective measures of upper extremity strength. The resultant simple linear regression equations based on rAmp significantly predicted both the grip force ($F(1,72)=8.57$, $R^2=0.09$, $p<0.01$) and lateral pinch grip strength ($F(1,72)=4.93$, $R^2=0.05$, $p=0.03$).

### *Other TMS and neurophysiology findings:*

An MEP response was observed in 29 subjects (76.32%) upon stimulation of both hemispheres. Absent MEP responses occurred in 7 subjects (18.42%). Two subjects (5.26%) had MEP responses from only one cortical hemisphere. Among those with absent MEPs 5 subjects (71.43%) had associated bulbar symptoms. Bulbar symptoms accompanied a significantly lower ALSFRSr score (p<0.01, 95% CI: 1.74, 8.24), MEP amplitude (p<0.01, 95% CI: 0.56, 2.37) and rAmp (p=0.04, 95% CI: 1.14, 42.02), compared to subjects without bulbar symptoms. Bulbar symptoms were also accompanied by a significantly higher motor threshold (p<0.01, 95% CI: -22.93, -3.54). No significant differences in CMAP amplitudes were observed between subjects with prominent bulbar symptoms and those without. A significant linear relationship was observed between CMCT, motor threshold and rAmp. The resultant simple linear regression equations using rAmp significantly predicted the CMCT ($F(1,74)=113.70$, $R^2=0.60$, $p<0.01$) and the motor threshold ($F(1,74)=145.3$, $R^2=0.66$, $p<0.01$). Both CMCT and motor threshold showed inverse linear relationship with rAmp.

***Predicting disease severity using neurophysiological measures:***

Using simple linear regression, we found that changes in rAmp was a significant predictor of disease severity measurements using ALSFRSr ($F(1,72)=19.35$, $R^2=0.20$, $p<0.01$) and AALS ($F(1,72)=11.03$, $R^2=0.12$, $p<0.01$). Simple linear regression models using the motor threshold and CMCT also significantly predicted both ALSFRSr ($F(1,72)=13.5$, $R^2=0.15$, $p<0.01$), ($F(1,72)=19.49$, $R^2=0.20$, $p<0.01$) and AALS ($F(1,72)=6.70$, $R^2=0.07$, $p=0.01$), ($F(1,72)=14.46$, $R^2=0.16$, $p<0.01$)) disease severity scores respectively. A summary of all the linear correlations between ALSFRSr scores and TMS parameters are shown in **figure 2a.**

Combining the TMS parameters together resulted in a significant multiple linear regression equation ($F(6,67)=4.35$, $R^2=0.22$, $p<0.01$). But, this MLR model constrained to the TMS data, needed improvement due to: 1) only accounting for a small proportion of the variability in ALSFRSr disease severity scores and 2) the floor effect of the prediction model. Therefore, to improve the prediction model, we combined both clinical and neurophysiological data. A correlation heat map (**figure 2b**) and the islands of statistically significant correlations plot (**figure 2c**), helped estimate the relative importance of each variable and build the MLR model for ALSFRSr prediction (**see supplementary methods**). The resultant MLR model accounted for a very high variance and significantly predicted the ALSFRSr scores ($F(21,46)=21.37$, $R^2=0.86$, $p<0.01$). But, this prediction model was less reliable due to: 1) the failure to identify an accurate relationship between ALSFRSr score and several of the predictor variables and 2) poor precision for the predicted values.

Therefore, the pooled dataset collected in this study was used to build a random forest based supervised ML model. Each variable used in this ML model had a unique significance in helping predict the disease severity (**see supplementary methods**). Since the variance accounted for by the MLR and ML prediction models were nearly 90%, a sample size of 38 subjects gave a statistical power of 99.96% for our conclusion that the MLR and ML based prediction models described here were significantly useful for disease prediction in ALS. Also, a subsequent double-blinded testing showed no significant differences between predicted and observed values for both MLR and ML models, with the ML model registering increased accuracy and precision, over the MLR model.

Significantly, the random forest based supervised ML model was capable of predicting disease severity with an overall accuracy of >97% and a precision of >99% (**table 2, figure 3a,b**). Other notable findings were: 1) the local weighted smoothening for the ALSFRSr - triple stimulation test data demonstrated an initial flattened disease progression slope; 2) for the corresponding degree of UMN dysfunction, both the regression lines and the local weighted smoothening lines for the ML prediction values closely followed those for the observed ALSFRSr values *(figure 3c)*.

### *Discussion:*

Transcranial magnetic stimulation (TMS) follows Faraday-Maxwell's laws of electrodynamics (*18*). A brief duration electrical current due to a rapidly varying magnetic field is induced in the cortical interneurons. The TMS recordings are often highly variable, due to the nature of how MEPs are generated. The triple stimulation technique (TST) was developed to reduce some of the inherent variability with TMS. The amplitude ratio of TST is representative of the fraction of the functional network of neurons in the corticospinal tract(*19*). The central motor conduction time (CMCT) quantifies the loss of fastest conducting neurons in the corticospinal tract network(*20*) and despite its test-retest variability, the longitudinal changes in CMCT has been shown to be a predictor of disease severity(*21*). Here, we demonstrate for the first time the role of cross-sectional measures of TST amplitude and area ratios, CMCT and motor threshold TMS parameters to be independently useful in predicting the degree of clinical impairment in ALS. We also successfully tested the ML prediction model for ALS disease severity.

Our modeling of the decaying magnetic field strength and the primary electric field induced on the surface of the brain, showed a significantly focal electric E field while using a figure-of-eight coil (*figure 1a,b*). A more focal field increased the accuracy and precision of the TMS recordings by selective activation of ADM. Another important requirement for accurate TST recordings was the nature of the TMS pulse. Usually single pulse protocols rely on a monophasic stimulation pulse. Since TST is dependent on maximal activation of the motor cortex neurons, our use of a biphasic pulse was important (***supplementary methods***). Improvement of TMS parameters based on physical properties of the brain stimulation coil, produced accurate TMS and TST recordings.

An important clinical finding is the strong concordance between measures of dexterity and TMS parameters. Clinical studies that measure corticomotoneuronal change usually include dexterity measures that involve timed repetitive fine-motor tasks. The degree of corticomotoneuronal dysfunction as measured by TST scores significantly predicted timed fine-motor repetitive tasks like block-and-board test and the Purdue peg board tests. We also found that hand weakness was also significantly influenced by corticomotoneuronal dysfunction. We also observed a significant correlation between CMAP amplitude and all four of the TMS parameters. Similar to previous studies, we noticed an inverse relation between distal CMAP amplitude and motor threshold (*22*). These findings together are suggestive of a possible influence of corticomotoneuronal dysfunction on motor units.

An exception to this inverse relation between the CMAP amplitude and the motor threshold was observed among subjects with bulbar symptoms. Bulbar symptoms usually accompany in-excitable motor cortex(*23*). This finding was consistent with our observation of significantly increased motor threshold, lower MEP amplitudes and lower TST amplitude ratios, despite robust CMAPs that accompanied bulbar symptoms. Since the neurophysiological effect of TMS is mediated predominantly through recruitment of the population of interneurons, these findings suggest an underlying upper motor neuron pathophysiology causing bulbar symptoms. These findings also suggest a differential involvement of motor cortex neuronal population in different phenotypes of ALS and are worth further exploration.

We also observed that an initial flattening of the ALSFRSr slope occurred among subjects with early corticomotoneuronal involvement. This flattening of the ALSFRSr slope could be explained in the context of 'e pluribus unum' model analysis of ensembles of accumulators(*24*), which found that firing rates are largely invariant with respect to ensemble size if the accumulators share at least modestly correlated accumulation rates. The subset of neurons with properties that can be considered outliers compared to the rest did not significantly alter the network properties. This model along with the neurophysiological findings of our study can explain some of the underlying motor cortical changes associated with ALS.

The initial flat slope of ALSFRSr in relation to the TMS parameters can be viewed as a state where the overall functional motor network characteristics undergo slower alterations compared to exponential loss in ensemble accumulators, an outcome due to the compensatory nature of the motor networks. Only when the neuronal loss exceeds a very large value, is a significantly observable influence in disease severity and neurophysiological changes recorded. These findings of an initial slow progression of the disease are also consistent with recent observations of cortical changes as the primary event in the disease process (*25*). The presence of a relatively stable curve of early disease progression compared to a greater functional loss of corticomotoneurons, followed by a rapid deterioration in clinical functions helps explain the non-linear disease progression pattern in ALS(*26*) and needs further investigation.

As an isolated clinical tool, TST is extremely useful due to its small range of normal values. For simple stratification as normal or abnormal findings, the TST recordings are easier to interpret than other TMS parameters. But this simple stratification alone had very little significance in predicting disease severity. Despite its small range for normal values and a significant correlation with disease severity measures, the TST scores were only moderately strong predictors of disease severity. This was applicable for all other individual TMS measures used in this study. However, both clinical and neurophysiological measures, when pooled together using random forest based supervised ML, the ALS disease severity prediction was improved with over 97% overall accuracy, 99% precision and accounting for nearly 90% of the variance in ALSFRSr scores.

Both the MLR and ML models were significant predictors of ALSFRSr scores. This indicates the robustness of the predictor variables used in both MLR and ML models. But, there is scope for improvement in accuracy, precision and ability to estimate the variations in disease progression. There have been previous proposals for combining ML and cortical excitability parameters (*27*). Combining additional TMS measures like the cortical silent period, input-output curve and paired pulse stimulation protocols might help add greater predictive power to the ML model. Also, neurophysiology measures and imaging measures have shown definite promise in identifying ALS disease severity (*28*). Therefore the usefulness of measures like motor unit number estimation and neuroimaging data including magnetic resonance (MR) spectroscopy, susceptibility weighted imaging and diffusion tensor imaging for improvement of the ML model, also need to be evaluated.

In previous studies, the utility of TST as a measure to predict structural abnormalities in ALS (*29*) and improve diagnostic accuracy of ALS has been demonstrated (*30*). In our study, we demonstrate for the first time that cross sectional TST measures are moderately strong, significant predictors of disease severity on their own. Using ML, this predictive power could be significantly boosted. The TST amplitude and area ratios were identified as important variables for disease severity prediction. Combining various clinical and neurophysiological measures with TST scores significantly improved the ability of the ML model to identify small changes in ALS disease severity. As a neurophysiology procedure, TST is no more time consuming or painful than a proximal erb's point stimulation and information from this procedure is easy to interpret. In conclusion, quantifying corticomotoneuronal dysfunction using upper extremity TST is extremely useful and should be included in a TMS study of ALS subjects. The random forest based cross-sectional ALSFRSr score prediction technique utilizing TST is a potential future biomarker for ALS disease severity.

*Figures and tables;*

| Description | Status |
|---|---|
| **Random forest regression** | |
| Number of trees | 2000 |
| No. of variables tried at each split | 7 |
| Mean of squared residuals | 4.28 |
| Variance in ALSFRr scores explained | 88.47% |
| **Multiple linear regression** | |
| Number of variables in the model | 21 |
| Mean of squared residuals | 5.23 |
| Variance in ALSFRr scores explained (adjusted $R^2$) | 86.1% |
| Significance level ($p$) | <0.01 |

*Table 1:* Summary of the characteristics of supervised random forest based machine learning and multiple linear regression models for ALSFRSr score prediction.

| Type of the model | | Precision | SD | Accuracy | SD |
|---|---|---|---|---|---|
| **Random forest regression** | Overall | 99.42% | 0.65% | 97.56% | 3.70% |
| | Blinded data | 99.22% | 0.43% | 92.06% | 6.52% |
| **Multiple linear regression** | Overall | 96.95% | 3.76% | 95.39% | 6.38% |
| | Blinded data | 97.53% | 3.67% | 86.62% | 10.96% |

*Table 2:* Comparison of mean absolute error (MAE) and root mean squared error (RMSE) for machine learning and the multiple linear regression models.

(a) 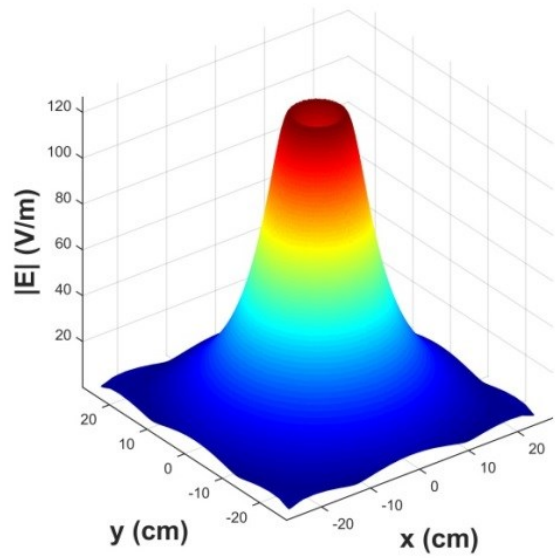 (b) 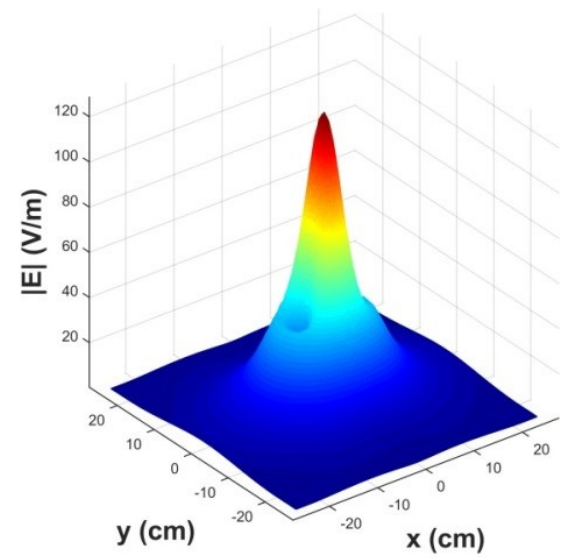

*Figure 1:* Computational modeling of E fields: *(a)* predicted E field norm 40 mm from the center of a circular coil, *(b)* predicted E field norm 40 mm from the center of a figure-of-eight coil.

*(a)*

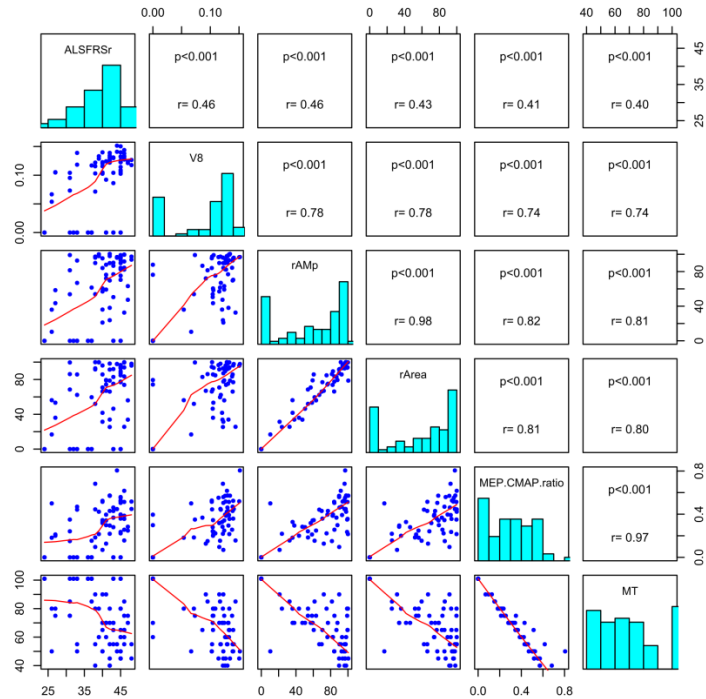

*(b)* *(c)*

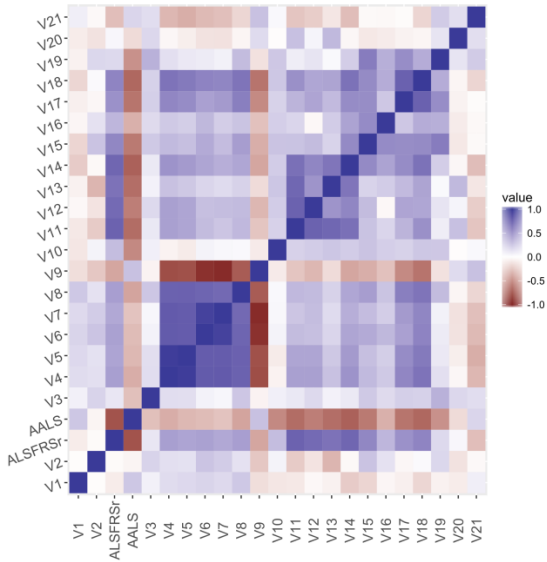 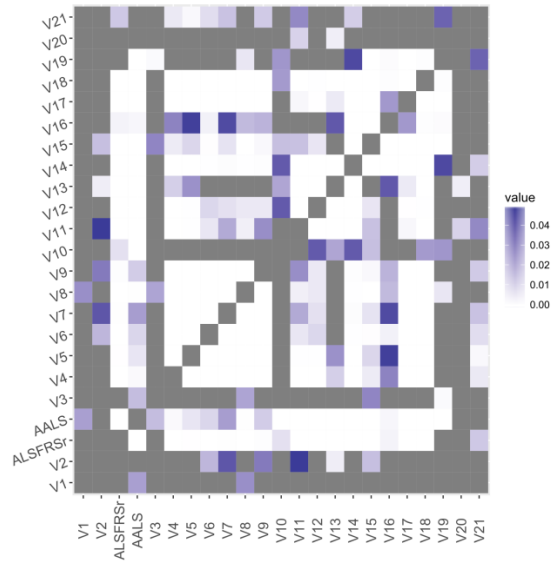

***Figure 2:*** Summary of the correlations observed between ALS disease severity and different clinical/neurophysiological measures: ***(a)*** matrix plot summarizing the significant (p<0.05) linear correlations between various TMS parameters, and ALS disease severities as measured using ALSFRSr, ***(b)*** heat-map plot summarizes the strength of correlations between all the modeling variables, ***(c)*** the islands of statistical significance plot highlights the significant (p<0.05) linear correlations between all the modeling variables.

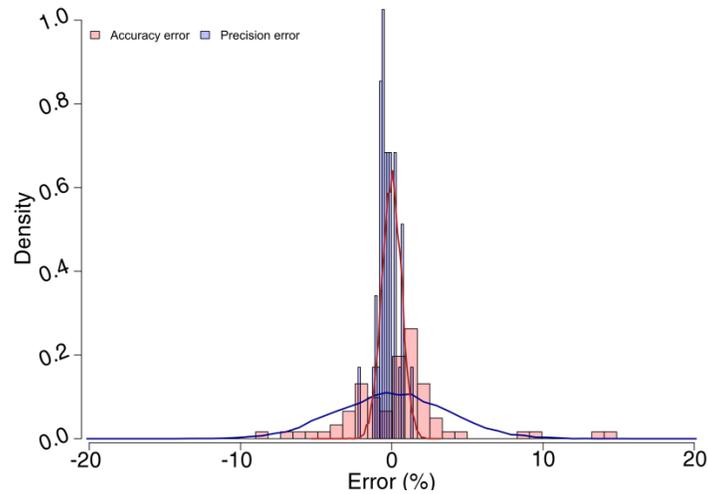
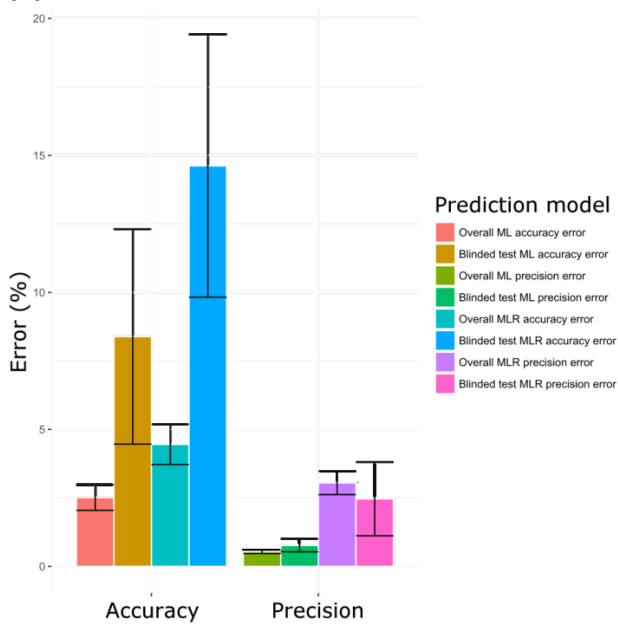
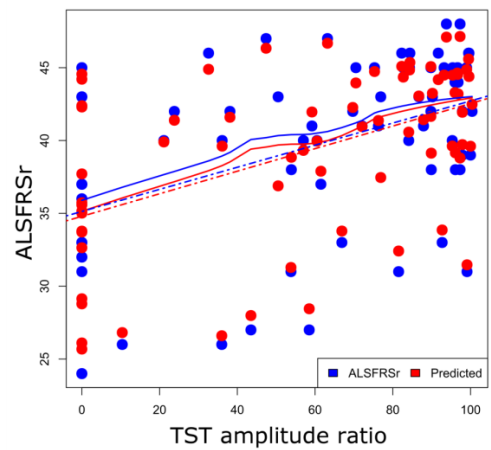

***Figure 3:*** Summary plots of the accuracy and precision of the machine learning (ML) disease severity for ALS prediction model: *(a)* histogram density plot of the accuracy and precision errors for the machine learning (ML) model and their corresponding normal distributions, *(b)* overall accuracy and precision errors for the multiple linear regression (MLR) and ML models compared against the blinded-test accuracy and precision errors for both models, *(c)* scatter plot comparison of TST amplitude ratio versus the observed (blue) and the ML predicted (red) ALSFRSr scores, with an overlay of simple linear regression (dotted) and local weighted smoothening (solid) lines.

***Competing financial interests:***
   None of the authors or immediate families have any competing financial relationships.

***Author contributions:***
***Rahul Remanan, M.B.B.S.***
   The author made substantial contributions to the development of this research project, including the transcranial magnetic stimulation procedures, neurophysiology assessments, clinical assessments, statistical analyses and development of machine learning prediction toolkit. He is the corresponding author of this article, and led the development of both the main text and the supplementary methods sections of this article.

*Viktor Sukhotskiy, Ph.D. graduate student.*
The author made substantial contributions to this research project, specifically in the development of computational modeling of the transcranial magnetic stimulations (TMS). He also made significant contributions to the development of the main text and supplementary methods sections of this article, highlighting the details of this computational simulation model of TMS.

*Mona Shahbazi, N.P.*
The author made substantial contributions to the development of the research project, especially those sections involving clinical assessments. Her contributions to this article include development of sections detailing the clinical assessments and descriptions of the nature of the disease process in the main text and in the supplementary methods sections.

*Edward P. Furlani, Ph.D.*
The author made substantial contributions to the development of the research project, specifically in the development of computational modeling of the transcranial magnetic stimulations (TMS). His role was pivotal in overseeing the development of this precise computational model. He also made significant contributions to the development of the main text and the supplementary methods sections of this article, highlighting the details of computational simulation model of TMS.

*Dale J. Lange, M.D.*
The author's contributions to this research project include the development of transcranial magnetic stimulation procedures, clinical assessments and statistical analyses. He had a pivotal role in overseeing the development of the neurophysiology and clinical assessment toolkits used in this study. He also made significant contributions to the development of the main text and supplementary methods of this article, especially those highlighting the clinical assessments and the nature of the disease process. He had a significant role in conceptual development of both the main text and the supplementary methods sections of this research article.

*Materials & Correspondence:*
All correspondences and material requests should be addressed to Rahul Remanan. The author can be reached at: remananr@hss.edu.

**Supplementary methods:**

*Magnetic field strength modeling:*
An understanding of physical mechanisms and properties of TMS is important in reducing measurement variability (*1*). Since the E field produced by the TMS coil(s) plays a critical role in this process, it is instructive to examine its spatial distribution for traditional coil configurations. To this end, we developed 3D computational models to investigate and compare the E field distributions generated by two different TMS coils, a circular coil (MagVenture C-100) and a butterfly, or "figure-of-eight" coil: the MagVenture MC-B70 model.

The **B** and **E** fields generated by the coils were computed using the Comsol Multiphysics AC/DC module with the Magnetic Fields (mef) physics interface on Microsoft Windows, version 5.2(*2*). This program solves the Maxwell system of equations [1]-[8] in the frequency-domain:

$$\nabla \times \mathbf{H} = \mathbf{J} \qquad [1]$$

$$\mathbf{B} = \nabla \times \mathbf{A} \qquad [2]$$

$$\mathbf{E} = -\frac{\partial \mathbf{A}}{\partial t} - \nabla \cdot V \qquad [3]$$

$$\mathbf{J} = \sigma \mathbf{E} + \frac{\partial \mathbf{D}}{\partial t} + \mathbf{J}_e \qquad [4]$$

$$\nabla \cdot \mathbf{J} = 0 \qquad [5]$$

$$\mathbf{D} = \varepsilon_0 \varepsilon_r \mathbf{E} \qquad [6]$$

$$\mathbf{B} = \mu_0 \mu_r \mathbf{H} \qquad [7]$$

$$\mathbf{n_2} \cdot (\mathbf{J_1} - \mathbf{J_2}) = 0 \qquad [8]$$

where, **E** is the electric field intensity, **D** is the electric displacement, **H** is the magnetic field intensity, **B** is the magnetic flux density, **J** is the total current density, **A** is the magnetic vector potential, $\varepsilon_0$ is the permittivity of vacuum, $\varepsilon_r$ is the relative permittivity of a medium, $\mu_0$ is the permeability of vacuum, $\mu_r$ is the relative permeability, $\sigma$ is the electrical conductivity, **J$_e$** is the applied current density, i.e. the current density imposed in the coils, V is the electric scalar potential and **n$_2$** is the outward normal from medium 2 at interfaces between two media (**J$_1$** and **J$_2$** are respectively the current densities of medium 1 and 2).

The Comsol Multiphysics software solves eqs. [1]-[8] numerically using the finite element method. The inputs to the computational models include the coil energization **J$_e$**, the material domains with corresponding boundaries, the material properties (magnetic, electric and dielectric) and appropriate physical boundary conditions. The coils' energization parameters were: a biphasic sinusoidal current with a peak magnitude of 690.7 A per turn and a frequency of 3571 Hz, which was modeled by extending a 280 μs sinusoidal biphasic pulse into a steady-state signal. The biphasic (full-sine) pulse is detailed in the MagPro user guide (Pg. 18). The coil current peak magnitude was derived from the peak di/dt value (155 A/μs) displayed by the MagPro R-30 during actual usage of the coils. The models were implemented using rectangular prism computational domains (CDs), wherein the TMS coil was surrounded by air (***figures S1a,b***). Three-dimensional meshes of tetrahedral linear elements of second order were used to tessellate the entire CD. The dimensions of the CD were increased and the meshes were refined until a stable solution and convergence errors lower than 1% were achieved. The details of the computational models are summarized in ***table S1***. The physical dimension data for the C-100 and MC-B70 coil material domains were taken from the coil vendor (*3, 4*).

Simulations were carried out using the linear Flexible Generalized Minimum Residual (FGMRES) iterative solver with 10,000 maximum number of iterations. This solver avoided out of memory problems, provided a fast convergence and computing robustness. A fully coupled approach using an automatic Newton method was chosen with a minimum damping factor of $1 \times 10^{-4}$ since it can ensure convergence effectiveness. These parameters were chosen because the initial conditions are well-known wherein all electromagnetic quantities are set to zero. The models were computed on a Precision® T7610 workstation (Dell Technologies Inc., Round Rock, TX, USA) with a 24 core Xeon® E5-2697 v2 processor (Intel Corp., Santa Clara, CA, USA) and using 64 GB DDR3 memory. The analysis for each coil took approximately 20 minutes to complete.

### *Screening and subject selection:*

A total of 57 subjects were screened from a database within a neuromuscular clinic population. From this screening pool of 57 subjects, 38 were included in the motor neuron disease (MND) category and underwent TMS, registering a screen failure rate of 33.33%. Among those 19 subjects who were screen failures 9 (47.37%) did not meet the diagnostic criteria, 7 (36.84%) were cancellations or voluntary withdrawal from the research study, 2 (10.53%) were due to the severity of the disease preventing them to undergo the procedure and 1 (5.26%) due to the discomfort associated with the procedure.

### *Recordings of the CMAP, erb's point, F-wave and MEP responses:*

The supramaximal distal compound muscle action potential (CMAP) and erb's point responses were required to perform TST. These responses were determined using an incremental stimulation intensity technique. The threshold surface electrode current required to elicit a ≥75µV amplitude response at 100µV sensitivity window from the ADM, was determined. The threshold response intensity was increased to four fold and a second set of responses were recorded. The corresponding responses were considered as the supramaximal distal CMAP and erb's point recordings. The stimulation intensity was increased by 5% to identify the plateauing of supramaximal response amplitudes. Among subjects in whom plateauing of the responses could not be elicited, stimulation intensity was increased by 5% intervals until a plateauing was observed. The corresponding responses were identified as the supramaximal distal CMAP and erb's recordings, among those subjects.

A band-pass setting of 2 Hz - 10 kHz without 60Hz filtering was used for all the TST acquisitions(*5*). To measure the F-wave latency, a set of 10 supramaximal stimulation responses were recorded. The late responses were recorded at sensitivity windows of 100 µV, with a bandpass setting of 2Hz – 3kHz, and no 60Hz filtering. The late response waveforms were averaged and the earliest late response was used to calculate the F-wave latency. The CMCT was calculated using the equation [9],

$$\mathbf{CMCT} = MEP_{latecny} - \frac{F-wave_{latency} + Distal\ CMAP_{latency} - 1}{2} \qquad [9]$$

where, F-wave latency was used to compute the peripheral nerve conduction delay that occurred between the anterior horn of the spinal cord and the ADM(*6*).

All study related recordings were performed in a specially designed room to minimize electrical interference. All recordings were made while the subject was sitting upright. The side from where the responses were recorded was suitably stabilized. The elbow joint was flexed at ~90° angle. The hands were in a supinated position. For MEP recordings, the initial placement of the stimulating coil was in such a way that the point of maximal magnetic field strength was along an imaginary point on the scalp, 4-5cm lateral to the vertex. The coil's longitudinal axis was in the anterior-posterior direction. The horizontal axis was along 45 degrees to an imaginary line connecting the vertex and tragus. The point of optimal stimulation for the region of the motor cortex that controls the fifth digit was identified by measuring the ADM responses and adjusting the positioning of the coil accordingly. Contralateral facilitation of the same muscle groups was performed in some subjects to optimize the MEP recordings. During facilitation, an auditory feedback was used to maintain a muscle contraction of ~20-30% of the maximal effort.

*Multiple linear regression model:*

Due to the inherent challenges associated with neuroscience data, both MLR models and supervised ML, have their own benefits and pitfalls in pattern predictions (*7, 8*). Therefore, testing both sets of models was important. First MLR model used only the TMS parameters. The resultant linear eq. [10],

$$\mathbf{ALSFRSr} = \sum_{i}^{n} \alpha_i x_i + 20.82 \qquad [10]$$

where, $\alpha_i$ is the coefficient for the variable $x_i$ and $n$ is the total number of variables in the predictor model (*table S3*). Due to the limitations in the TMS model a more detailed MLR model using a combination of both clinical and neurophysiological data was developed for ALSFRSr score prediction. The resultant linear eq. [11],

$$\mathbf{ALSFRSr} = \sum_{i}^{n} \beta_i x_i + 40.27 \qquad [11]$$

where, $\beta_i$ is the coefficient for the variable $x_i$ and $n$ is the total number of variables in the predictor model (*table S4*). This MLR model was compared against the supervised ML based model that used the same predictor variables.

*Machine learning model for ALSFRSr prediction:*

The random forest based supervised machine learning prediction model has been successfully tested for ALSFRSr score predictions from longitudinal data (*9, 10*). Using this supervised machine learning technique, a cross sectional ALSFRSr score predictor was developed as part of this study. A total of 21 variables were used in the ensemble training by developing predictor vectors for the random forest regression (*table S2a,b*). The relative importance of each of the predictors was estimated using mean increase in error and mean increase in node purity (*figure S2a,b*). Using a total of 1280 random objects generated with Marsenne twister pseudorandom generator that can be reproduced (*11*), a forest of 2000 trees was built and tested (*figure S2c*). Random number generators are implemented at the architectural level for newer generation microprocessors (*12, 13*). But, to ensure a wider compatibility and better reproducibility of the prediction code, a pseudorandom number generator implementation was used here. The machine learning computations can be performed with minimal computational overhead. The prediction models were tested in OptiPlex® 9010 workstation (Dell Technologies Inc., Round Rock, TX, USA) with a 4 core Core® i7-3770 processor (Intel Corp., Santa Clara, CA, USA), 4GB DDR3 memory. The necessary code for running the prediction model and de-identified datasets are available on GitHub (*14*).

*Statistical tests and testing of the prediction models:*

The descriptive statistics and within group analysis were done using the code published in the GitHub repository. The outputs from the statistical testing of the dataset are also included here (*14*). The two-sample Welch's unequal variance t-test was used to identify significant differences within groups. Using the initial recruitment of 35 subjects, the MLR and ML models were developed, followed by an addition of three subjects to the study pool; two females and one male. Two had features of bulbar predominant motor neuron disease. These three subjects were used to build a blinded-test data by blinding the observed ALSFRSr scores from the prediction scores. These de-identified blinded datasets are also included in the GitHub repository (*14*). One of the subjects in the blinded-test dataset underwent the same list of procedures four years prior. Since the goal of this study was to identify cross-sectional neurophysiology and clinical data for ALSFRSr score prediction, this particular dataset was treated as a unique cross-sectional data-point. Addition of this subject also helped in identifying any possible pitfalls in ML model while handling longitudinal data instead of cross-sectional data. Various prediction models were tested against this blinded-test data.

The accuracy, precision and corresponding standard deviations (SD) for both the final prediction models using MLR and ML were calculated using eqs. [12, 13, 14],

$$\Delta_{accuracy} = \sum_{i=1}^{n} \left( \left| 1 - \frac{ALSFRSr_{Observed} - ALSFRSr_{predicted}}{ALSFRSr_{Observed}} \right| \right) \times 100 \qquad [12]$$

$$\Delta_{precision} = \sum_{i=1}^{n} \left( \left| 1 - \frac{ALSFRSr_{predicted.right} - ALSFRSr_{predicted.left}}{48} \right| \right) \times 100 \qquad [13]$$

$$\Delta_{SD} = \sqrt{\frac{1}{n} \sum_{i=1}^{n} (\Delta_i)} \times 100 \qquad [14]$$

where, *n* is the total number of samples used in the prediction model. Both the MLR and ML prediction models computed unique ALSFRSr scores that corresponded to the right and left sides of each subject. These values were computed using a combination of both side dependent and side independent measures. The resultant values were used to compute the precision for both MLR and ML models. Since the ALSFRSr scale values ranges from 0 to 48, the maximum possible value for the scale was the denominator for computing percentage precision.

The random forest regression is based on random sampling of a vector, such that the corresponding tree predictor assumes a numerical value. Therefore the regression results are state dependent on the random attributes of the computational environment running the ML model. This results in slight differences in outputs for each separate run. The error analyses reported here are based on a randomly chosen single run of the prediction model based on the parameters described above. The outputs from the successful execution of the code that were used for the analyses, are also included in the GitHub repository(*14*).

*Supplementary figures and tables:*

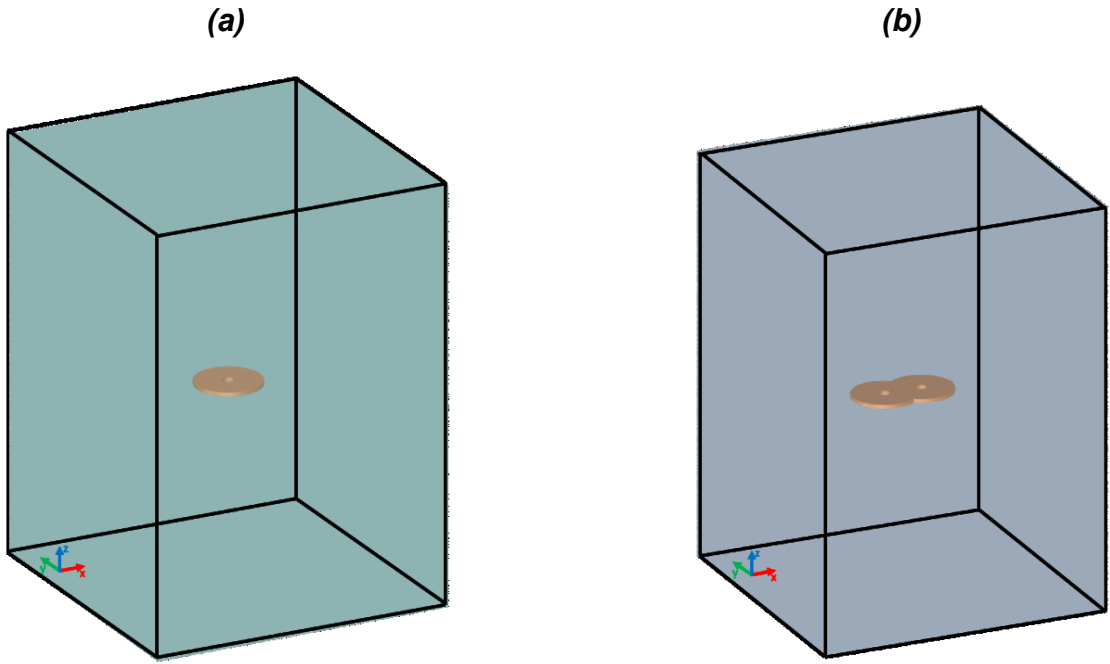

***Figure S1:*** Computational domain for numerical analysis of B and E fields showing TMS coil(s) in air: ***(a)*** single coil, ***(b)*** figure-of-eight coil.

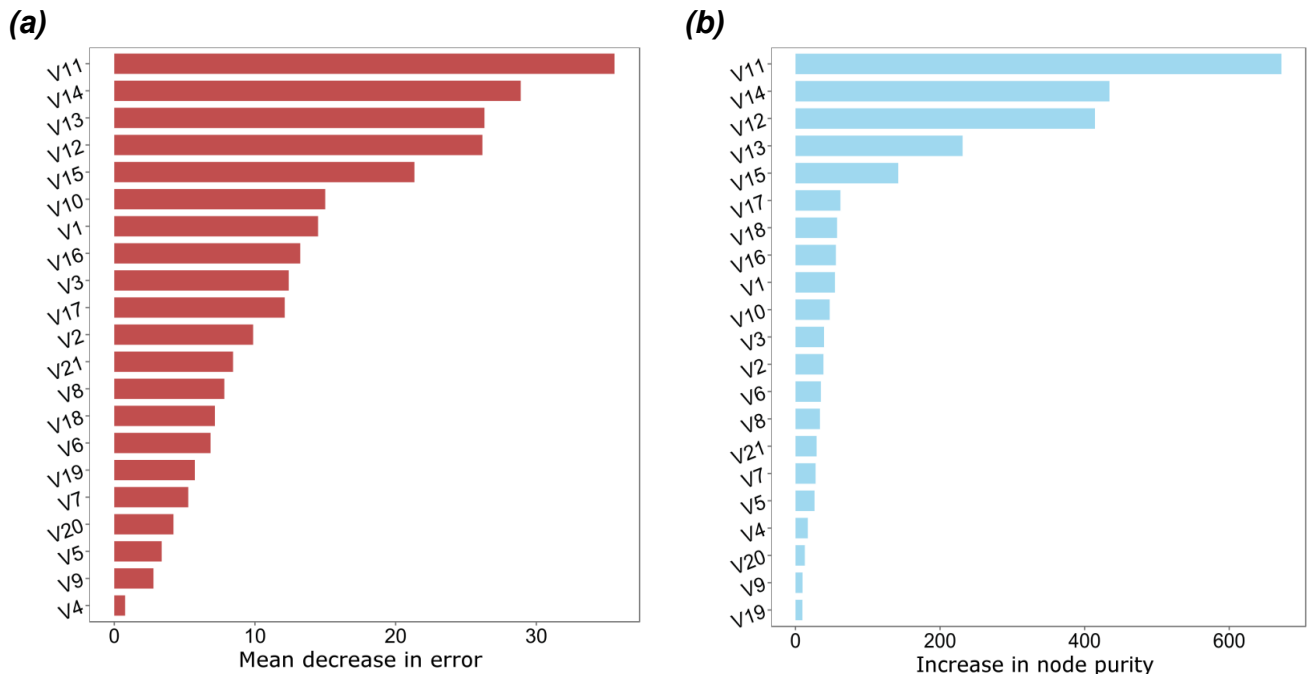
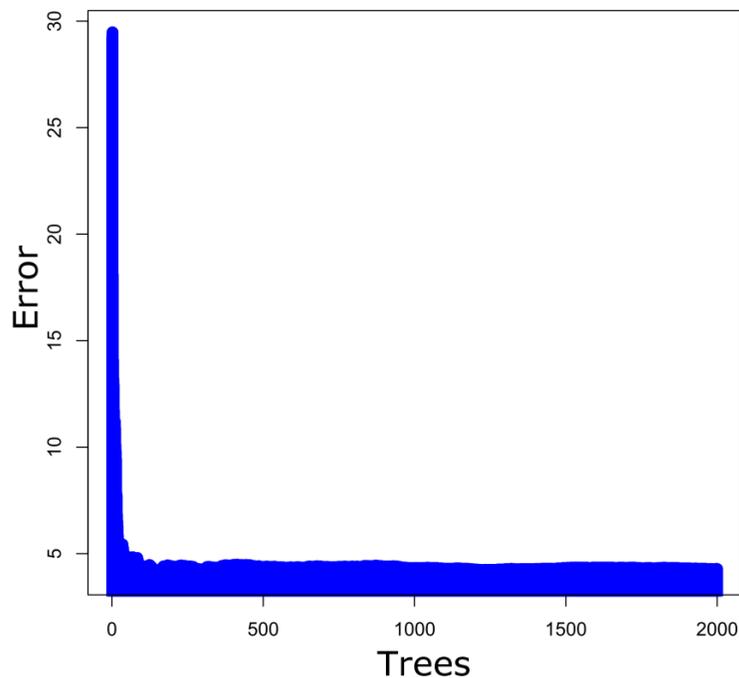

*Figure S2:* Relative importance of each predictor vectors used in the random forest based machine learning model for ALSFRSr score prediction: *(a)* mean decrease in error plot, *(b)* the expected error rate using Gini impurity index during the random splits training of the prediction model, *(c)* relative error rate in predicting ALSFRSr compared against the number of prediction trees used to build the ML model.

| Simulation type | Relative electric permittivity | Electric conductivity (S/m) | Relative magnetic permeability | Dimensions | | Total number of tetrahedral and boundary mesh elements | Degrees-of-freedom |
|---|---|---|---|---|---|---|---|
| **C-100** | 1 | $6.0 \times 10^7$ | 1 | **Coil** | Coil height: 6 mm | 455733 | 4135240 |
| | | | | | Outside diameter: 110 mm | | |
| | | | | | Inside diameter: 20 mm | | |
| | | | | | Cross sectional area (wire): 19.29 mm$^2$ | | |
| | | | | | Length (wire): 3 m | | |
| | | | | **Air block** | Width (x): 500 mm | | |
| | | | | | Depth (y): 500 mm | | |
| | | | | | Height (z): 750 mm | | |
| **MC-B70** | 1 | $6.0 \times 10^7$ | 1 | **Coil** | Coil height: 6 mm | 505332 | 4612009 |
| | | | | | Outside diameter: 97 mm | | |
| | | | | | Inside diameter: 27 mm | | |
| | | | | | Cross sectional area of wire: 21 mm$^2$ | | |
| | | | | | Length (wire): 3 m | | |
| | | | | **Air block** | Width (x): 500 mm | | |
| | | | | | Depth (y): 500 mm | | |
| | | | | | Height (z): 750 mm | | |

***Table S1***: Details of dimensions of computational domain, dielectric and magnetic properties of materials and media used for simulation of the C-100 and MC-B70 TMS coils.

| Vector | Description of the vector | Mean decrease in error | Increase in node purity |
|---|---|---|---|
| V1 | Age in years | 14.49 | 54.60 |
| V2 | Body mass index | 9.87 | 38.64 |
| V3 | McGill single item quality of life score | 12.40 | 39.48 |
| V4 | TST amplitude ratio as percentage | 0.77 | 17.04 |
| V5 | TST area ratio as percentage | 3.37 | 26.37 |
| V6 | MEP:CMAP amplitude ratio | 6.84 | 35.13 |
| V7 | MEP: proximal stimulation amplitude ratio | 5.26 | 27.80 |
| V8 | Inverse of central motor conduction time measured in milliseconds | 7.82 | 33.82 |
| V9 | Motor threshold as a percentage of the maximal output of the TMS stimulator | 2.79 | 9.76 |
| V10 | Forced vital capacity as percentage of the predicted | 15.00 | 47.31 |
| V11 | Inverse of the time to complete walking 20ft measured in seconds | 35.56 | 672.24 |
| V12 | Inverse of the time to standup from a chair measured in seconds | 26.17 | 414.26 |
| V13 | Inverse of the time to climb up and down 4 steps of stairs measured in seconds | 26.31 | 231.18 |
| V14 | Inverse of the time to propel a wheelchair 20ft measured in seconds | 28.89 | 434.36 |
| V15 | Grip force testing using Jamar dynamometer in kg | 21.34 | 142.17 |
| V16 | Lateral pinch grip in kg | 13.22 | 55.89 |
| V17 | Timed Purdue peg board test | 12.12 | 62.14 |
| V18 | Timed block and board test | 7.15 | 57.52 |
| V19 | Manual muscle strength testing of the upper extremity using MRC scale | 5.73 | 9.58 |
| V20 | Manual muscle strength testing of the lower extremity using MRC scale | 4.21 | 12.94 |
| V21 | Duration of the disease in years | 8.44 | 29.20 |
| V22 | Gender (0=male, 1=female) | | |
| V23 | Visual analogue scale | | |
| V24 | Presence or absence of bulbar symptoms (0=no, 1=yes) | | |
| V25 | Race (1=Caucasian, 2=Black/African American, 3=Asian, 4=Hispanic) | **Unused vectors for the ML model** | |
| V26 | Handedness (0=left,1=right) | | |
| V27 | Smoking status (0=no, 1=yes) | | |
| V28 | Distal compound muscle action potential (CMAP) amplitude (µV) | | |
| V29 | Proximal stimulation response amplitude (µV) | | |
| V30 | Motor evoked potential amplitude (µV) | | |

**Table S2:** Description of vectors and relative importance of each one when used for the development of random forest based supervised machine learning prediction of ALSFRSr scores (*see figure S2a,b*).

| Variable | Coefficients | | t value | Significance level | |
|---|---|---|---|---|---|
| | B | Standard error | | | |
| (Constant) | 20.8195 | 14.2171 | 1.464 | 0.148 | |
| V4 | 0.14831 | 0.09111 | 1.628 | 0.108 | |
| V5 | -0.1174 | 0.08995 | -1.305 | 0.196 | |
| V6 | 11.937 | 8.5384 | 1.398 | 0.167 | |
| V7 | -0.5043 | 14.9013 | -0.034 | 0.973 | . |
| V8 | 28.6969 | 20.4667 | 1.402 | 0.165 | |
| V9 | 0.13729 | 0.14158 | 0.97 | 0.336 | |
| Significance codes: 0 '***' 0.001 '**' 0.01 '*' 0.05 '.' 0.1 ' ' 1 | | | | | |

*Table S3:* Summary of the multiple linear regression model using TMS measures as independent variables and the ALSFRSr score as the dependent variable.

| Variable | Coefficients | | t value | Significance level | |
|---|---|---|---|---|---|
| | B | Standard error | | | |
| (Constant) | 40.27 | 13.74 | 2.93 | 0.01 | ** |
| V1 | 0.05 | 0.04 | 1.18 | 0.24 | |
| V2 | 0.29 | 0.11 | 2.74 | 0.01 | ** |
| V3 | -0.24 | 0.15 | -1.61 | 0.12 | |
| V4 | -0.01 | 0.04 | -0.26 | 0.80 | |
| V5 | -0.03 | 0.04 | -0.67 | 0.51 | |
| V6 | -5.77 | 4.85 | -1.19 | 0.24 | |
| V7 | 13.15 | 7.27 | 1.81 | 0.08 | . |
| V8 | 18.87 | 11.56 | 1.63 | 0.11 | |
| V9 | 0.07 | 0.08 | 0.80 | 0.43 | |
| V10 | 0.00 | 0.02 | 0.11 | 0.91 | |
| V11 | 45.31 | 13.30 | 3.41 | 0.00 | ** |
| V12 | 1.86 | 2.59 | 0.72 | 0.48 | |
| V13 | 25.03 | 7.65 | 3.27 | 0.00 | ** |
| V14 | -13.87 | 12.48 | -1.11 | 0.27 | |
| V15 | 0.09 | 0.02 | 4.66 | 0.00 | *** |
| V16 | 0.01 | 0.04 | 0.25 | 0.80 | |
| V17 | 0.09 | 0.08 | 1.11 | 0.27 | |
| V18 | 0.02 | 0.04 | 0.59 | 0.56 | |
| V19 | -0.79 | 0.24 | -3.21 | 0.00 | ** |
| V20 | -0.22 | 0.11 | -2.03 | 0.05 | * |
| V21 | -0.18 | 0.10 | -1.82 | 0.08 | . |
| Significance codes: 0 '***' 0.001 '**' 0.01 '*' 0.05 '.' 0.1 ' ' 1 | | | | | |

*Table S4:* Summary of the multiple linear regression model using a combination of clinical and neurophysiology data as independent variables and the ALSFRSr score as the dependent variable.